\documentclass{nature}
\usepackage{amsmath,amssymb}
\usepackage{mathtools}
\usepackage{url}
\usepackage{hyperref}
\usepackage{lineno}

\usepackage{amssymb}
\usepackage[dvipsnames]{xcolor}

\title{Acceleration of 1I/`Oumuamua from radiolytically produced H$_2$ in H$_2$O ice}

\author{Jennifer B. Bergner$^{1,2}$ \& Darryl Z Seligman $^3$}

\begin{document}

\maketitle

\vspace{-0.2in}
\begin{affiliations}
 \item University of California, Berkeley, Department of Chemistry, Berkeley, CA 94720, USA; email: jbergner@berkeley.edu; ORCID: 0000-0002-8716-0482
  \item University of Chicago Department of the Geophysical Sciences, Chicago, IL 60637, USA
 \item Department of Astronomy and Carl Sagan Institute, Cornell University, 122 Sciences Drive, Ithaca, NY, 14853, USA; email: dzs9@cornell.edu; ORCID: 0000-0002-0726-6480
\end{affiliations}

\begin{abstract}
In 2017, 1I/`Oumuamua was identified as the first known interstellar object in the Solar System\cite{Williams17}.  Although typical cometary activity tracers were not detected\cite{Meech2017,Fitzsimmons2017,Jewitt2017, Ye2017, Trilling2018}, `Oumuamua exhibited a significant non-gravitational acceleration\cite{Micheli2018}.  To date there is no explanation that can reconcile these constraints\cite{Jewitt2022}.  Due to energetic considerations, outgassing of hyper-volatile molecules is favored over heavier volatiles like H$_2$O and CO$_2$\cite{Seligman2020}.  However, there are are theoretical and/or observational inconsistencies\cite{Levine2021b} with existing models invoking the sublimation of pure H$_2$ \cite{Seligman2020}, N$_2$, \cite{Desch20211i} and CO \cite{Seligman2021}.  Non-outgassing explanations require fine-tuned formation mechanisms and/or unrealistic progenitor production rates \cite{Micheli2018,moro2019fractal,Sekanina19b,Luu2020}.
Here we report that the acceleration of `Oumuamua is due to the release of entrapped molecular hydrogen which formed through energetic processing of an H$_2$O-rich icy body.  In this model, `Oumuamua began as an icy planetesimal that was irradiated at low temperatures by cosmic rays during its interstellar journey, and experienced warming during its passage through the Solar System.  This explanation is supported by a large body of experimental work showing that H$_2$ is efficiently and generically produced from H$_2$O ice processing, and that the entrapped H$_2$ is released over a broad range of temperatures during annealing of the amorphous water matrix\cite{Bar-Nun1985,Sandford1993,Watanabe2000,Grieves2005,Zheng2006,Zheng2006b,Zheng2007}.  We show that this mechanism can explain many of `Oumuamua's peculiar properties without fine-tuning.  This provides further support\cite{Fitzsimmons2017} that `Oumuamua originated as a planetesimal relic broadly similar to Solar System comets.

\end{abstract}

After the identification of `Oumuamua as an interstellar object, extensive follow-up studies revealed it to host a featureless red reflection spectrum \cite{Fitzsimmons2017, Meech2017, Jewitt2017,Ye2017}, no detectable dust or gas coma \cite{Meech2017,Jewitt2017,Trilling2018}, an extreme 6:6:1 geometry \cite{Meech2017, Jewitt2017,Mashchenko2019}, and a low inbound galactic velocity dispersion implying a young $\lesssim$100~Myr age \cite{Mamajek2017,Hallatt2020,Jewitt2022}. Considering these constraints, together with the observed non-gravitational acceleration\cite{Micheli2018}, we propose that `Oumuamua originated as an H$_2$O-rich icy planetesimal that was ejected from its formation system and underwent energetic processing in the cold interstellar medium. This bombardment by cosmic rays and high energy photons produced entrapped H$_2$ that was released upon thermal annealing of the H$_2$O matrix during its passage through the Solar System.  This model does not require a fine-tuned formation scenario, and reconciles the observations that (i) the surface properties of `Oumuamua are consistent with some Solar System comets, (ii) spectroscopically active gases were not detected, and (iii) the body nonetheless exhibited non-gravitational acceleration.  Note that H$_2$ would not be detectable from the spectroscopic observations of `Oumuamua at optical, centimeter, and IR wavelengths which were used to provide upper limits on the production rates of other gases\cite{Ye2017, Park2018, Trilling2018}.

The processing of H$_2$O ice in astrophysically relevant conditions has been extensively explored with laboratory experiments.  H$_2$ formation from H$_2$O processing (or D$_2$ from D$_2$O) has been demonstrated for various energy sources (energetic particles, UV photons, electrons), water structures (crystalline and amorphous), and compositions (water ice that is both pure  and doped with other volatiles)\cite{Bar-Nun1985,Sandford1993,Watanabe2000,Grieves2005,Zheng2006,Zheng2006b,Zheng2007}. Of relevance to the highly porous structure of comet-like bodies, H$_2$ production has been shown to increase along with ice porosity \cite{Grieves2005,Zheng2007}. The majority of H$_2$ produced from in situ ice processing at low temperatures ($\sim$10 K) remains trapped within the water matrix until the ice is heated to temperatures ranging from 15 to 140 K\cite{Sandford1993, Zheng2006}.  

Non-gravitational acceleration of some form is required to match the 207 astrometric observations of `Oumuamua collected when it was between 1.2--2.8 au \cite{Micheli2018,Seligman2021}.  The best fit to the non-gravitational acceleration is primarily in the anti-solar direction with an $r^{-1}$ or $r^{-2}$ radial dependency.  During this time (and specifically during the first 6 October nights when high temporal cadence photometric data were obtained) the lightcurve was used to infer object dimensions of 115$\times$111$\times$19 meters, though the size is degenerate with the geometric albedo\cite{Mashchenko2019} (the quoted dimensions assume an albedo of 0.1).  Before this time, it is unknown whether `Oumuamua experienced non-gravitational acceleration and what its shape or size may have been. 

H$_2$/H$_2$O yields as high as 35\% have been measured in experiments of H$_2$O ice processing\cite{Sandford1993}.  H$_2$/H$_2$O yields of tens of percent are predicted for the top few meters of a cometary body exposed to galactic cosmic rays (GCRs)\cite{Maggiolo2020}.  Indeed, GCRs are the most relevant energy source for this scenario since they can penetrate to depths of tens of meters into an icy body\cite{Gronoff2020}.  Moreover, in the solar system, the H/(C+O) ratio in volatile cometary material is $\sim$1.6, compared to a ratio of $\sim$0.5 in refractory organic material\cite{Derenne2010, Rubin2019}.  This implies that a majority of the original H content can be lost during chemical processing of volatile ice mixtures.  Based on the above lines of evidence, it is reasonable to assume that H$_2$/H$_2$O yields in the low tens of percent can be achieved for the top several meters of an irradiated comet-like body.  In our model we test uniform fixed H$_2$/H$_2$O ratios of 0.3 and 0.4 throughout the body, corresponding to conversion yields H$_2$(final)/H$_2$O(initial) of $\sim$20--30\%.

In order to be a feasible accelerant, H$_2$ outgassed from the surface layers of `Oumuamua must provide sufficient force to explain the observed non-grativational acceleration.  To test this model, we calculate the number of H$_2$ molecules in an outgassing shell on the surface of the ellipsoidal body ($N_\mathrm{H2, shell}$), and compare this to the total number of molecules required to explain the non-gravitational acceleration ($N_\mathrm{H2, accel}$) (Methods; Equation \ref{eq:metric}).  When $N_\mathrm{H2, shell}$ \big{/} $N_\mathrm{H2, accel}$ $>$1, the model is feasible. This fraction depends on several parameters which are unconstrained for `Oumuamua: the albedo (and corresponding size) of the body, the outgassing depth, the temperature of the outgassing H$_2$, the geometry of outflowing gas (e.g.~collimated vs.~isotropic), the H$_2$/H$_2$O ratio, and the water/dust mass ratio.  We therefore explore a range of values reasonable for a comet-like body, described in the Methods section.  Figure 1 demonstrates that the feasibility criterion is satisfied for a wide range of the expected parameter space in both optimistic and conservative cases.

\begin{figure}
\begin{center}
       \includegraphics[width=\linewidth]{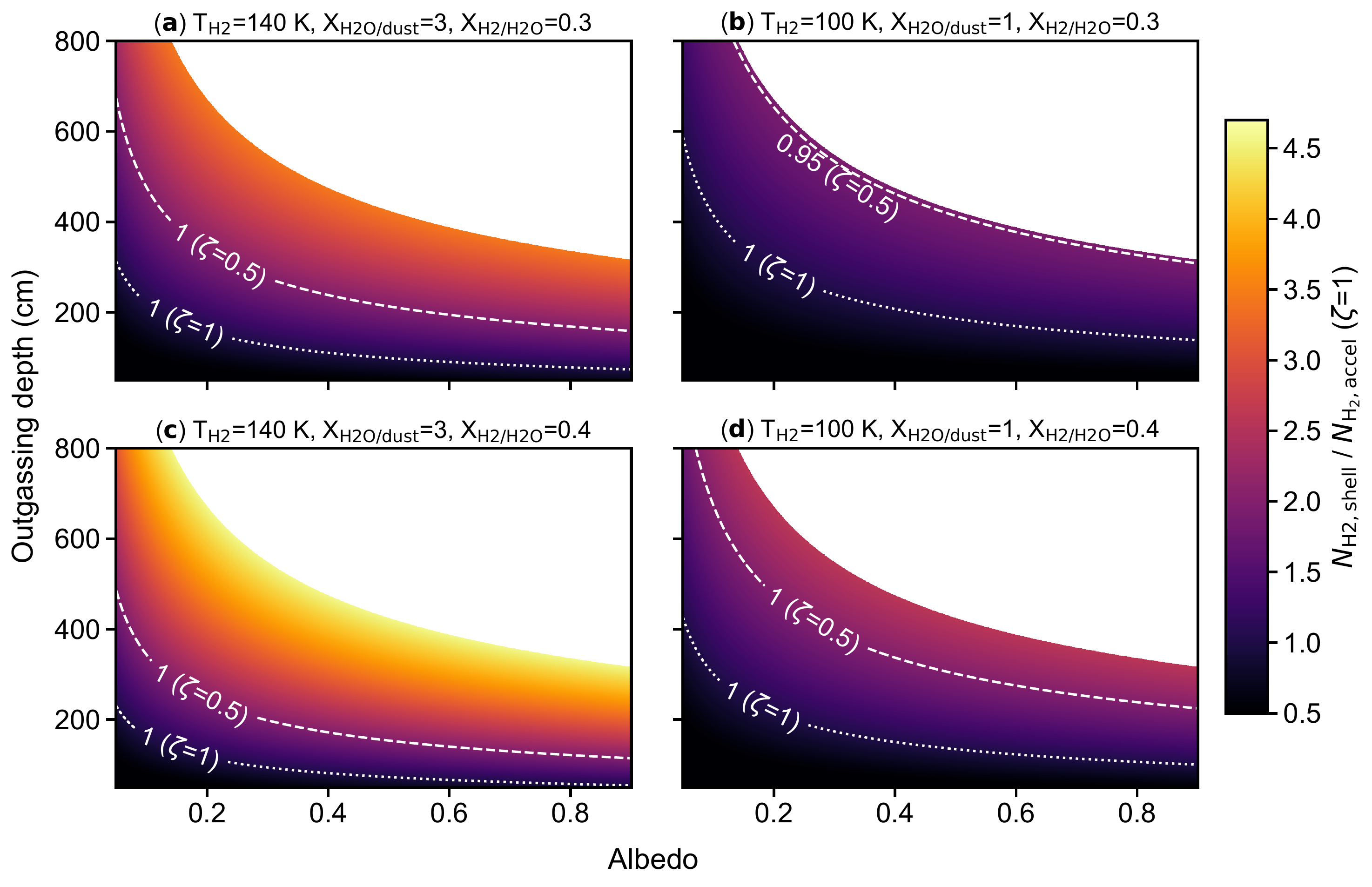}
    \caption{The model feasibility for a range of possible parameter space.  The model is feasible when $N_\mathrm{H2, shell}$/$N_\mathrm{H2, accel}$ $>$1, i.e.~the number of H$_2$ molecules in an outgassing shell is sufficient to explain the observed non-gravitational acceleration between 1.2--2.8 au.  Each panel shows this ratio for a range of geometric albedos and outgassing depths.  The white regions represent unphysical combinations of albedo and outgassing depth ($d > r_c$).  Different assumptions for the outflow temperature, H$_2$O/dust mass ratio, and H$_2$/H$_2$O number density ratio are shown in Panels A-D.  Lines show where the ratio is 1 for the case of a collimated outflow (dotted, $\zeta$=1) and an entirely isotropic hemispherical outflow (dashed, $\zeta$=0.5).  Regions above these contours represent allowable parameter space for a given set of assumptions.  In panel B the dashed line corresponds to a ratio of 0.95, the maximum reached for that $\zeta$=0.5 scenario.  Colormaps correspond to the $\zeta$=1 case for each panel.} \label{fig:h2_budget}
\end{center}
\end{figure}

We next perform thermal modeling of `Oumuamua (Methods) to determine whether the object could produce H$_2$ from the depths implied by Figure 1.  Experiments have demonstrated that H$_2$ is released from within an amorphous H$_2$O ice matrix between 15--140 K, first due to sublimation of H$_2$ from channels with access to the surface ($<$30 K), and then due to annealing of the H$_2$O ice which causes micropores and macropores to collapse\cite{Laufer1987,Grieves2005,Zheng2007}.  The fractions released within different temperature regimes have been estimated at 2/3 from 15--30 K, 1/9 from 30--80 K, and 2/9 from 80--140 K\cite{Bar-Nun1988}, though this may differ in ices of cometary thickness compared to the experiments.  Based on these experiments, we expect H$_2$ outgassing to be active for comet layers that are warmed to $\sim$15--140 K.  Figure 2 shows the simulated radial temperature profile of `Oumuamua over the course of the observations.  We show the temperature profile for two end member values of the thermal conductivity \cite{Gundlach2012}.  For a low thermal conductivity, temperatures of 15--140 K are reached between 50--250 cm below the surface, while for a higher thermal conductivity the necessary temperatures are reached down to $\sim$8 meters.  It is possible that additional heating may be produced from the exothermicity of crystallization or annealing of the water matrix\cite{Prialnik2022}. Upon comparison with Figure 1, it is clear that the nucleus can be warmed to sufficient depths to explain the necessary extent of H$_2$ outgassing (e.g.~200--400 cm for moderate albedos).

\begin{figure} 
\begin{center}
       \includegraphics[scale=0.43,angle=0]{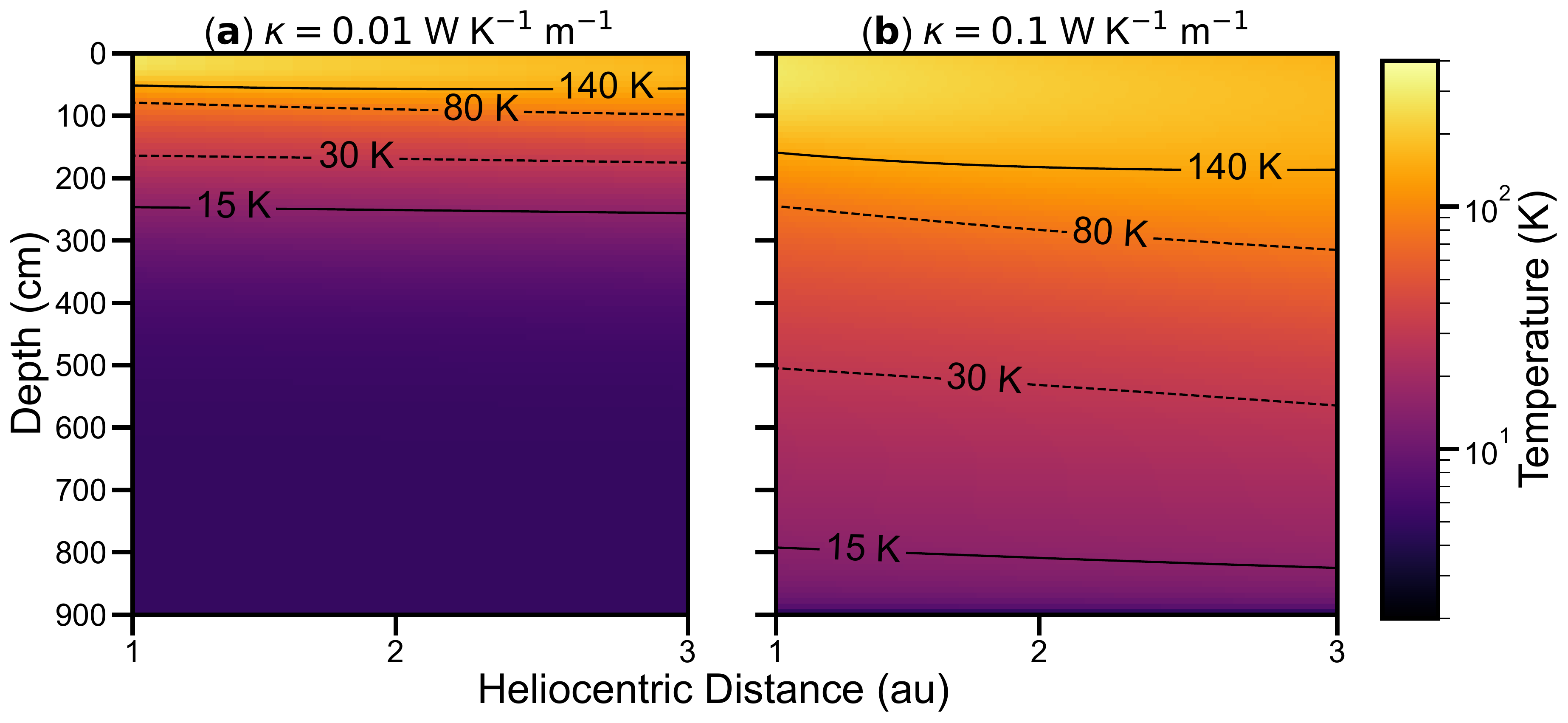}
    \caption{Thermal model of `Oumuamua during the observational arc where non-gravitational acceleration is required.  We adopt a bulk density $\rho_{Bulk}=0.5$ g cm$^{-3}$, specific heat capacity of $c_P=2000$ J kg$^{-1}$ K and thermal conductivity $\kappa=10^{-2}$ W K$^{-1}$ m (panel A) and $\kappa=10^{-1}$ W K$^{-1}$ m (panel B).  H$_2$ outgassing is expected for temperatures of $\sim$15--140 K.}\label{fig:thermal} 
\end{center}
\end{figure}

Our thermal models predict that surface temperatures greater than the water sublimation temperature ($\sim$150 K) can be reached interior to 3 au.  While it has been shown that H$_2$O alone could not provide sufficient non-gravitational acceleration due to energetic constraints \cite{Sekanina2019, Seligman2020}, spectroscopic observations did not rule out H$_2$O outgassing.  Indeed, upper limits from radio lines of OH (a dissociation product of H$_2$O) with the Green Bank Telescope correspond to an upper limit on the water mass loss rate of $\dot{M} \le$ 30 kg s$^{-1}$ at 1.8 au\cite{Park2018,Jewitt2022}, which is considerably higher than the production that would be needed to power the non-gravitational acceleration \cite{Micheli2018}.  If H$_2$O sublimation was active, this would lower the required yields of H$_2$ needed to explain the non-gravitational acceleration.  Based on the energetic constraints from \cite{Seligman2020}, H$_2$O sublimation could explain at most $\sim$50\% of the observed non-gravitational acceleration.  Alternatively, H$_2$O may not have been present near the surface due to the formation of a low-conductivity regolith during interstellar radiation processing\cite{Jewitt2017, Fitzsimmons2017, Seligman2018}.  In any case, we have demonstrated that radiolytically produced H$_2$ could provide the non-gravitational acceleration of `Oumuamua despite the ambiguity regarding H$_2$O production.

Our modeling approach is necessarily simple owing to a dearth of quantitative constraints.  Additional laboratory efforts are needed to enable a more sophisticated treatment.  In particular, we stress the need to robustly establish the yields of H$_2$ from different types and doses of energetic processing, both for pure H$_2$O ice and comet-like volatile ice mixtures.  This also includes measuring the H$_2$ fraction that escapes in different temperature regimes for different ice thicknesses.  Better descriptions of the energetics and kinetics of amorphous H$_2$O ice annealing are also needed.

A remaining obstacle to any outgassing-based explanation for `Oumuamua's non-gravitational acceleration is the upper limit on the reported (and model-dependant) $\mu$m sized dust production ($\dot{M} \le$ 2$\times 10^{-4}$ -- 2$\times 10^{-3}$ kg s$^{-1}$), which is lower than expected based on active solar system comets \cite{Jewitt2017,Meech2017}.  Micheli et al.~\cite{Micheli2018} posit that the dust from `Oumuamua could have been released as larger grains ($>$100$\mu$m), and also note that small dust activity is not universal in solar system comets\cite{Fink2009}. For example, the well studied comet 2P/Encke is notoriously underabundant in small dust particles \cite{Newburn1985}, though this may be due to thermal processing within the inner solar system.  We expand upon these arguments by considering $\mu$m sized dust that is initially on the surface of the object or trapped within the subsurface ice matrix.  It has been demonstrated that interactions with ambient gas in the interstellar medium preferentially remove small dust grains from the surface of long-period comets and interstellar comets \cite{stern1990}. Furthermore, `Oumuamua is considerably smaller than any active comet detected to date in the solar system\cite{Jewitt2022}.  It is possible that smaller objects with weaker self gravity are less efficient at retaining surface dust particles.   Indeed, the lack of known small Solar System comets could reflect an observational bias due to intrinsically lower dust production rates.  In the scenario proposed here, we further note that H$_2$ is less massive than other volatiles associated with cometary activity (e.g. 2 vs.~28 m$_\mathrm{H}$ for H$_2$ vs.~CO). Therefore, H$_2$ production will provide considerably less momentum for entraining dust either from the surface or within the ice matrix.  Similarly, it is is plausible that H$_2$O annealing will release entrapped H$_2$ while the primordial dust remains bound in the ice matrix.  Improved constraints on the efficiency of small dust entrainment in the outgassing of different hypervolatiles are important not only for `Oumuamua but for interpreting cometary activity more generally. 

The model proposed here is generic and implies that all icy bodies exposed to sufficient high energy radiation should exhibit some degree of H$_2$O conversion to H$_2$. This process should be particularly efficient for Oort Cloud comets that are not shielded from cosmic rays by the heliosphere.  It is important to note the small size of `Oumuamua, which is at least an order of magnitude smaller than solar system comets with `classic' activity (e.g.~spectroscopically active volatiles and/or small dust; see Figure 5 in \cite{Jewitt2022}).  Because ice processing will be most efficient in the top few meters of the surface of a comet\cite{Maggiolo2020}, the outgassing of H$_2$ from a larger, more massive body should not be expected to produce a meaningful acceleration.  Therefore, it is likely that the detection of non-gravitational acceleration of `Oumuamua was possible owing to its small size.  

\begin{figure} 
\begin{center}
       \includegraphics[scale=0.43,angle=0]{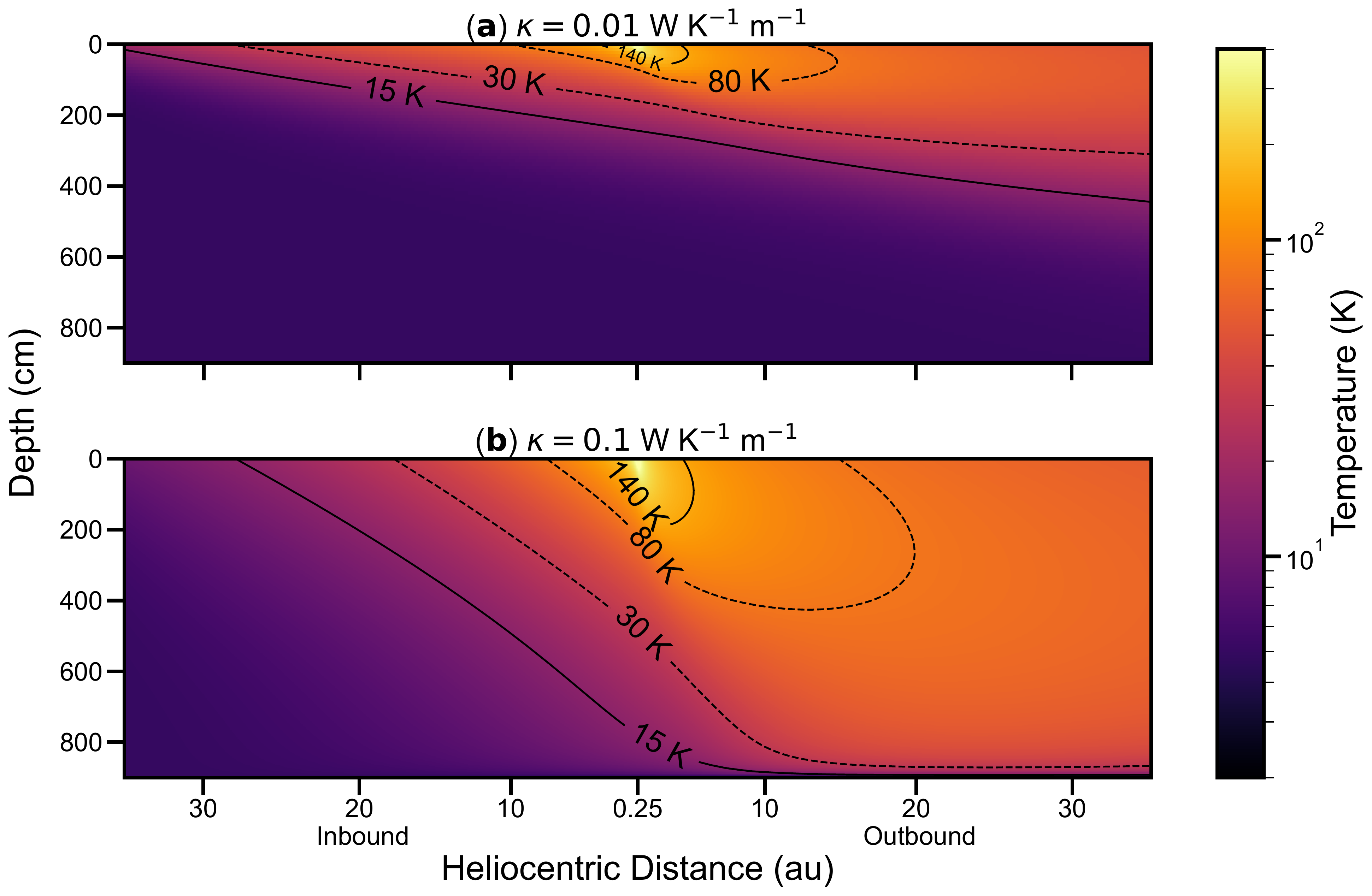}
    \caption{Thermal model of ‘Oumuamua over the course of the entire trajectory interior to the orbit of Neptune.  Adopted parameters are the same as in Figure 2.}\label{fig:prediction} 
\end{center}
\end{figure}

A prediction of our model is that small long-period and interstellar comets should exhibit non-gravitational accelerations analogous to `Oumuamua as they approach the Sun, even if they exhibit faint or non-detectable activity. Figure 3 shows the thermal profile of an object on the same trajectory as `Oumuamua to distances past Neptune.  Based on this model, H$_2$ outgassing from the surface layers of icy bodies may begin at large heliocentric distances (e.g.~outgassing down to 2 meters between 10--20 au).  Follow-up observations of small solar system bodies and interstellar objects discovered with the Rubin Observatory Legacy Survey of Space and Time (LSST) will therefore enable the proposed mechanism to be tested.  Future detections of small bodies with non-gravitational acceleration and faint coma could provide insights into the origins of `Oumuamua even though it has long departed the Solar System.

\begin{methods}

\subsection{H$_2$ budget}

We first calculate the total number of H$_2$ molecules needed to explain the non-gravitational acceleration between 1.2--2.8 au.  Following the fits of \cite{Micheli2018}, we assume the acceleration as a function of heliocentric distance $r$ is described by $a (r)$=4.92$\times$10$^{-6} \Big{(}r\big/\mathrm{1 au}\Big{)}^{-2}$ m s$^{-2}$.  The mass production rate $Q$ of sublimating molecules can be found using,
\begin{equation}
  M\,a\big(r\big)\, = \zeta Q v_\mathrm{H2},
\end{equation}
where $M$ is the mass of the body, $v_\mathrm{H2}$ is the velocity of the outgassing H$_2$, and $\zeta$ represents the fraction of outgassing momentum that is imparted in the anti-solar direction.  $v_\mathrm{H2}$ is found from $(\frac{8}{\pi}k_B T_\mathrm{H2}/m_\mathrm{H2})^{1/2}$, where $T_\mathrm{H2}$ is the temperature of outflowing H$_2$ and $m_\mathrm{H_2}$ is the H$_2$ mass.  As discussed in more detail subsequently, we assume $T_\mathrm{H2}$ is comparable to or lower than the surface temperature of the body, reflecting full or partial thermalization as H$_2$ escapes.  Assuming an ellipsoidal shape, we solve for the \textit{instantaneous} H$_2$ production rate needed to power the non-gravitational acceleration, $\dot{N}_\mathrm{H2, accel}$ (molecules s$^{-1}$):
\begin{equation}
  \dot{N}_\mathrm{H2, accel} = \,\bigg(\,\frac{4}{3}\pi r_a r_b r_c \rho_{Bulk}a\,\bigg)\,\bigg/\,\bigg(\zeta m_\mathrm{H2}v_\mathrm{H2}\,\bigg)\,,
\end{equation}
where $r_a$, $r_b$, and $r_c$ are the semi axes of the ellipsoid and $\rho_{Bulk}$ is the density.  The \textit{total} number of required H$_2$ molecules $N_\mathrm{H2, accel}$ is found by integrating $\dot{N}_\mathrm{H2, accel}$ over time during the trajectory.  Note that we only integrate over the portion of the trajectory during which astrometric measurements of the object were obtained, e.g.~between 1.2--2.8 au, denoted by times $t_i$ and $t_f$:

\begin{equation}\label{eq:NH2_accel}
    N_\mathrm{H2, accel} = \int_{t_i}^{t_f} \,\dot{N}_\mathrm{H2, accel} \big(\,\tau\,\big)\, {\rm d \tau}\, .
\end{equation}

To evaluate the feasibility of our hypothesis, we compare this with the number of H$_2$ molecules contained within a volumetric, ellipsoidal shell with a finite thickness $d$.  To do so we first solve for the volumetric density of H$_2$ (molecules cm$^{-3}$) within the body:
\begin{equation}\label{eq:volumetric_density}
    n_\mathrm{H2} = \frac{\rho_{Bulk} X_\mathrm{H2O/dust} X_\mathrm{H2/H2O}}{m_\mathrm{H2O}},
\end{equation}
where $X_\mathrm{H2O/dust}$ is the H$_2$O:dust mass ratio, $X_\mathrm{H2/H2O}$ is the H$_2$:H$_2$O number ratio, and $m_\mathrm{H2O}$ is the mass of H$_2$O.  The RHS of Equation \ref{eq:volumetric_density} is multiplied by the shell volume to calculate the  total number of molecules in the outgassing shell, $N_\mathrm{H2, shell}$,

\begin{equation}\label{eq:NH2Shell}
    N_\mathrm{H2, shell}=  X_\mathrm{H2O/dust}\,X_\mathrm{H2/H2O}\frac{4 \pi}{3}\,\bigg(\,\frac{\rho_{Bulk}}{m_\mathrm{H2O}}\,\bigg)\, \,\bigg(\, r_a r_b r_c -  (r_a-d) (r_b-d) (r_c-d)\,\bigg)\,.
\end{equation}
With this treatment, we implicitly assume that the H$_2$ content within an entire ellipsoidal shell should ultimately be available for outgassing.  At a given time we expect that outgassing will only occur from the illuminated surface; however, ‘Oumuamua was tumbling with a rotational period much shorter than the time in which its heliocentric distance changed substantially.  Therefore, over the trajectory the entire surface of the body should have been exposed to the Sun and subject to H$_2$ outgassing.

We adopt as our feasibility criterion the dimensionless quantity given by dividing Equation \ref{eq:NH2Shell} by Equation \ref{eq:NH2_accel}:

\begin{equation}\label{eq:metric}
\begin{split}
    N_\mathrm{H2, shell}\bigg/N_\mathrm{H2, accel}=   X_\mathrm{H2O/dust}\,X_\mathrm{H2/H2O}\,\bigg(\,\frac{m_\mathrm{H2}}{m_\mathrm{H2O}}\,\bigg)\, \,\bigg(\, 1 -  \frac{(r_a-d) (r_b-d) (r_c-d)}{r_a r_b r_c}\,\bigg)\\\,\,\bigg(\,v_\mathrm{H2} \zeta \bigg/\int_{t_i}^{t_f} \,a \big(\,\tau\,\big)\, {\rm d \tau}\,\bigg)\,.
    \end{split}
\end{equation}
When the fraction $N_\mathrm{H2, shell}$/$N_\mathrm{accel}$ $>$ 1, the scenario is feasible: the number of H$_2$ molecules in the shell is sufficient to explain the non-gravitational acceleration.  Note that this condition is related to the model feasibility starting around the time of detection (roughly 1 au post-perihelion).  As discussed subsequently, pre-discovery outgassing might require a larger $N_\mathrm{H2, shell}$.  In Equation \ref{eq:metric}, we assume that the velocity of the outgassing H$_2$ is independent of heliocentric distance.  $N_\mathrm{H2, shell}$/$N_\mathrm{accel}$ does not formally depend on $\rho_{Bulk}$.  We assume a constant $X_\mathrm{H2/H2O}$ of either 0.3 or 0.4, corresponding to an H$_2$ yield of 23--28\% of the original H$_2$O content, as described in the main text.  Other unconstrained parameters that influence $N_\mathrm{H2, shell}$/$N_\mathrm{H2, accel}$ are the outgassing depth, the size of the body, the outgassing temperature, the outflowing gas geometry (approximated by $\zeta$), and the mass ratio $X_\mathrm{H2O/dust}$.  The size of the body is degenerate with geometric albedo ($p$), and we explore a range of albedos from 0.05--0.9.  The axis lengths of the ellipsoid ($r_a$, $r_b$, and $r_c$) are found by rescaling the dimensions found for an albedo of 0.1 (115 x 111 x 19 meters) by a factor of $\sqrt{0.1/p}$ \cite{Mashchenko2019}. We also model a range of outgassing depths from 100--800 cm to explore the depth required to explain the magnitude of observed non-gravitational acceleration.  Note that for high albedos, the nominal outgassing depth can be larger than the minor axis $r_c$; these cases are unphysical and result in the blank region of parameter space shown in Figure 1.  

For $T_\mathrm{H2}$ and $X_\mathrm{H2O/dust}$, we evaluate two end member cases.  For the first, less conservative case we assume an outgassing temperature of 140 K and a H$_2$O/dust mass ratio of 3:1.  This temperature assumes that the escaping H$_2$ molecules are thermalized as they leave the surface.  While higher surface temperatures are possible (Figure 2), the water matrix will begin to sublimate somewhat above 140 K.  A water:dust mass ratio of 3:1 was found to be the best fit to an outgassing model of `Oumuamua by \cite{Micheli2018}.  For the second, more conservative model we assume a lower outgassing temperature of 100 K, representative of a scenario in which the escaping H$_2$ is not fully thermalized with the surface layers of the body; and a lower H$_2$O/dust mass ratio of 1:1.  The second model is conservative in that both the lower temperature and the lower $X_\mathrm{H2O/dust}$ result in a lower $N_\mathrm{H2, shell}$ for a given albedo and outgassing depth.

We also test two cases for the outflow geometry.  For a fully collimated outflow, all momentum from the outgassing H$_2$ is imparted in the anti-solar direction, corresponding to $\zeta$ = 1.  For an entirely isotropic hemispherical outflow, half of the momentum is anti-solar, corresponding to $\zeta$ = 0.5.  In reality, the outflow geometry is likely in between these cases.  Note that in addition to the parameters explored here, a momentum contribution from H$_2$O sublimation would further expand the allowable parameter space compared to what is shown in Figure 1.

For reference, we calculate the the H$_2$ mass production rate $Q$ at 1.25 au.  Assuming an outgassing temperature of 100 K and $\zeta$=0.75, the mass loss rate ranges from $\sim$10--800 g s$^{-1}$ for the high- and low-albedo extremes, respectively.  This is lower than previous estimates of the H$_2$O mass loss rate required to power the non-gravitational acceleration \cite{Micheli2018,Jewitt2022}, due to a combination of (i) a higher gas velocity due to the smaller mass of H$_2$ compared to H$_2$O, and (ii) different assumptions about the size of the body.

\subsection{Thermal model}

In this section we describe our numerical thermodynamic model of the interior temperature of `Oumuamau. This calculation is similar to the numerical experiments published by \cite{Fitzsimmons2017} and \cite{Seligman2018}. The absorbed solar flux, $\Phi_\odot$, along the trajectory is,
\begin{equation}\label{eq:Flux}
    \Phi_\odot(t) = \,\bigg(\,\frac{\xi\,L_\odot }{4\pi R(t)^2}\,\bigg)\,\,\bigg(\,1-p\,\bigg)\, \, ,
\end{equation}
where $p$ is the albedo, $R(t)$ is the time-dependent heliocentric distance and $\xi=0.25$ is the average projected surface area \cite{Meltzer1949}. We adopt $p$=0.1 as the fiducial albedo; the resulting temperature profile is only weakly sensitive to the choice of albedo, and a higher albedo for the thermal model does not affect our conclusions.  We solve the radial temperature profile by solving the heat conduction equation in cylindrical coordinates,

\begin{equation}\label{eq:CylHeatEq}
    \frac{dT}{dt}=\frac{\kappa}{\rho_{Bulk} c_P}\frac{1}{r}\frac{\partial}{\partial r}\,\bigg(\,r\frac{\partial T}{\partial r}\,\bigg) \, ,
\end{equation}
for the temperature $T$. In Equation \ref{eq:CylHeatEq}, $\kappa$ is the thermal conductivity,  $c_P$ is the specific heat capacity, and $r$ is the  depth. We implement an iterative scheme that uses a second order Newton-Raphson technique to solve for the temperature at the surface, T$_{\rm Surf}$.  This T$_{\rm Surf}$ must balance (i) the absorption of solar flux ($\Phi_\odot(t)$), (ii) the heating of the surface element material ( $c_P \rho_{Bulk} \Delta r \Delta T_{\rm Surf}$) (iii) the reradiation from the surface ($\epsilon \sigma T_{\rm Surf}^4$), and (iv) diffusion of heat  from the surface  into the interior ($\kappa\, \,{dT/dr}_{r=0}$). The outer boundary condition is given by, 

\begin{equation}\label{eq:OuterBC}
    \Phi_\odot(t) -\epsilon \sigma T_{\rm Surf}^4 - \kappa  \frac{dT}{dr}_{r=0}-c_P \rho_{Bulk} \Delta r \Delta T_{\rm Surf}=0\, .
\end{equation}
In Equation \ref{eq:OuterBC}, $\epsilon \sim0.95$ is the emissivity and $\sigma$ is the Stefan-Boltzmann constant.  Because the thermal properties of `Oumuamua are  unconstrained, we assume a bulk density $\rho_{Bulk}=0.5$ g cm$^{-3}$, typical of cometary nuclei \cite{britt2006}, and a specific heat capacity of $c_P=2000$ J kg$^{-1}$ K$^{-1}$ typical for cometary materials and ices.  We produce thermal models for $\kappa=10^{-2}$ W K$^{-1}$ m$^{-1}$ and $\kappa=10^{-1}$ W K$^{-1}$ m$^{-1}$ \cite{Steckloff2021}.

Our thermal model implies that H$_2$ outgassing may be possible well before perihelion. Specifically, temperatures $>15$ K are achieved down to depths of $\sim$2 and $\sim$8 meters prior to perihelion for thermal conductivities of $\kappa=10^{-2}$ and $\kappa=10^{-1}$ W K$^{-1}$ m, respectively (Figure 3).   This raises the question of whether it is sufficient to explain `Oumuamua's non-gravitational acceleration only during the portion of the trajectory from 1--3 au post-perihelion when astrometric measurements were obtained.  However, there are major challenges associated with projecting any model prior to the observable apparition.  It is not obvious that the non-gravitational acceleration, which was required to remove long-term trends from 76 days of astrometry\cite{Micheli2018}, can be extrapolated to the entire trajectory.  Other equally plausible scenarios exist, including that (i) the presence of a low-conductivity regolith delayed the propagation of warming until post-perihelion\cite{Fitzsimmons2017, Seligman2018}; or (ii) the body experienced ice sublimation around perihelion, leading to changes in its size and/or shape\cite{Seligman2020}, with the implication that the original reservoir of H$_2$ (and H$_2$O) available to power an outgassing recoil was larger than what would be inferred from the observations starting at 1.2 au post-perihelion.  Given these uncertainties, we only consider outgassing between $\sim$1--3 au when there are constraints on `Oumuamua’s dimensions and non-gravitational acceleration.  Still, we note that for some parameter space (Figure 1), the outgassing shell contains factors of a few more H$_2$ molecules than is necessary to explain the non-gravitational acceleration from 1--3 au. It therefore remains plausible that  H$_2$ outgassing was ongoing prior to perihelion. 

\end{methods}

\bibliography{sample}

\begin{thebibliography}{10}
\expandafter\ifx\csname url\endcsname\relax
  \def\url#1{\texttt{#1}}\fi
\expandafter\ifx\csname urlprefix\endcsname\relax\def\urlprefix{URL }\fi
\providecommand{\bibinfo}[2]{#2}
\providecommand{\eprint}[2][]{\url{#2}}

\bibitem{Williams17}
\bibinfo{author}{{Williams}, G.~V.} \emph{et~al.}
\newblock \bibinfo{title}{{Minor Planets 2017 SN\_33 and 2017 U1}}.
\newblock \emph{\bibinfo{journal}{Central Bureau Electronic Telegrams}}
  \textbf{\bibinfo{volume}{4450}}, \bibinfo{pages}{1} (\bibinfo{year}{2017}).

\bibitem{Meech2017}
\bibinfo{author}{{Meech}, K.~J.} \emph{et~al.}
\newblock \bibinfo{title}{{A brief visit from a red and extremely elongated
  interstellar asteroid}}.
\newblock \emph{\bibinfo{journal}{Nature}} \textbf{\bibinfo{volume}{552}},
  \bibinfo{pages}{378--381} (\bibinfo{year}{2017}).

\bibitem{Fitzsimmons2017}
\bibinfo{author}{{Fitzsimmons}, A.} \emph{et~al.}
\newblock \bibinfo{title}{{Spectroscopy and thermal modelling of the first
  interstellar object 1I/2017 U1 `Oumuamua}}.
\newblock \emph{\bibinfo{journal}{Nature Astronomy}}
  \textbf{\bibinfo{volume}{2}}, \bibinfo{pages}{133--137}
  (\bibinfo{year}{2018}).

\bibitem{Jewitt2017}
\bibinfo{author}{{Jewitt}, D.} \emph{et~al.}
\newblock \bibinfo{title}{{Interstellar Interloper 1I/2017 U1: Observations
  from the NOT and WIYN Telescopes}}.
\newblock \emph{\bibinfo{journal}{\apjl}} \textbf{\bibinfo{volume}{850}},
  \bibinfo{pages}{L36} (\bibinfo{year}{2017}).

\bibitem{Ye2017}
\bibinfo{author}{{Ye}, Q.-Z.}, \bibinfo{author}{{Zhang}, Q.},
  \bibinfo{author}{{Kelley}, M.~S.~P.} \& \bibinfo{author}{{Brown}, P.~G.}
\newblock \bibinfo{title}{{1I/2017 U1 ({\lsquo}Oumuamua) is Hot: Imaging,
  Spectroscopy, and Search of Meteor Activity}}.
\newblock \emph{\bibinfo{journal}{\apjl}} \textbf{\bibinfo{volume}{851}},
  \bibinfo{pages}{L5} (\bibinfo{year}{2017}).

\bibitem{Trilling2018}
\bibinfo{author}{{Trilling}, D.~E.} \emph{et~al.}
\newblock \bibinfo{title}{{Spitzer Observations of Interstellar Object
  1I/{\lsquo}Oumuamua}}.
\newblock \emph{\bibinfo{journal}{\aj}} \textbf{\bibinfo{volume}{156}},
  \bibinfo{pages}{261} (\bibinfo{year}{2018}).

\bibitem{Micheli2018}
\bibinfo{author}{{Micheli}, M.} \emph{et~al.}
\newblock \bibinfo{title}{{Non-gravitational acceleration in the trajectory of
  1I/2017 U1 ('Oumuamua)}}.
\newblock \emph{\bibinfo{journal}{Nature}} \textbf{\bibinfo{volume}{559}},
  \bibinfo{pages}{223--226} (\bibinfo{year}{2018}).

\bibitem{Jewitt2022}
\bibinfo{author}{{Jewitt}, D.} \& \bibinfo{author}{{Seligman}, D.~Z.}
\newblock \bibinfo{title}{{The Interstellar Interlopers}}.
\newblock \emph{\bibinfo{journal}{arXiv e-prints}}
  \bibinfo{pages}{arXiv:2209.08182} (\bibinfo{year}{2022}).

\bibitem{Seligman2020}
\bibinfo{author}{{Seligman}, D.} \& \bibinfo{author}{{Laughlin}, G.}
\newblock \bibinfo{title}{{Evidence that 1I/2017 U1 ('Oumuamua) was Composed of
  Molecular Hydrogen Ice}}.
\newblock \emph{\bibinfo{journal}{\apjl}} \textbf{\bibinfo{volume}{896}},
  \bibinfo{pages}{L8} (\bibinfo{year}{2020}).

\bibitem{Levine2021b}
\bibinfo{author}{{Levine}, W.~G.}, \bibinfo{author}{{Cabot}, S. H.~C.},
  \bibinfo{author}{{Seligman}, D.} \& \bibinfo{author}{{Laughlin}, G.}
\newblock \bibinfo{title}{{Constraints on the Occurrence of 'Oumuamua-Like
  Objects}}.
\newblock \emph{\bibinfo{journal}{\apj}} \textbf{\bibinfo{volume}{922}},
  \bibinfo{pages}{39} (\bibinfo{year}{2021}).

\bibitem{Desch20211i}
\bibinfo{author}{Desch, S.~J.} \& \bibinfo{author}{Jackson, A.~P.}
\newblock \bibinfo{title}{1i/‘oumuamua as an n2 ice fragment of an exo-pluto
  surface ii: Generation of n2 ice fragments and the origin of ‘oumuamua}.
\newblock \emph{\bibinfo{journal}{Journal of Geophysical Research: Planets}}
  \bibinfo{pages}{e2020JE006807} (\bibinfo{year}{2021}).

\bibitem{Seligman2021}
\bibinfo{author}{{Seligman}, D.~Z.}, \bibinfo{author}{{Levine}, W.~G.},
  \bibinfo{author}{{Cabot}, S. H.~C.}, \bibinfo{author}{{Laughlin}, G.} \&
  \bibinfo{author}{{Meech}, K.}
\newblock \bibinfo{title}{{On the Spin Dynamics of Elongated Minor Bodies with
  Applications to a Possible Solar System Analogue Composition for 'Oumuamua}}.
\newblock \emph{\bibinfo{journal}{\apj}} \textbf{\bibinfo{volume}{920}},
  \bibinfo{pages}{28} (\bibinfo{year}{2021}).

\bibitem{moro2019fractal}
\bibinfo{author}{Moro-Mart{\'\i}n, A.}
\newblock \bibinfo{title}{Could 1i/’oumuamua be an icy fractal aggregate?}
\newblock \emph{\bibinfo{journal}{\apjl}} \textbf{\bibinfo{volume}{872}},
  \bibinfo{pages}{L32} (\bibinfo{year}{2019}).

\bibitem{Sekanina19b}
\bibinfo{author}{{Sekanina}, Z.}
\newblock \bibinfo{title}{{1I/`Oumuamua and the Problem of Survival of Oort
  Cloud Comets Near the Sun}}.
\newblock \emph{\bibinfo{journal}{arXiv e-prints}}
  \bibinfo{pages}{arXiv:1903.06300} (\bibinfo{year}{2019}).

\bibitem{Luu2020}
\bibinfo{author}{{Luu}, J.~X.}, \bibinfo{author}{{Flekk{\o}y}, E.~G.} \&
  \bibinfo{author}{{Toussaint}, R.}
\newblock \bibinfo{title}{{'Oumuamua as a Cometary Fractal Aggregate: The
  ``Dust Bunny'' Model}}.
\newblock \emph{\bibinfo{journal}{\apjl}} \textbf{\bibinfo{volume}{900}},
  \bibinfo{pages}{L22} (\bibinfo{year}{2020}).

\bibitem{Bar-Nun1985}
\bibinfo{author}{{Bar-Nun}, A.}, \bibinfo{author}{{Herman}, G.},
  \bibinfo{author}{{Rappaport}, M.~L.} \& \bibinfo{author}{{Mekler}, Y.}
\newblock \bibinfo{title}{{Ejection of H $_{2}$O, O $_{2}$, H $_{2}$ and H from
  water ice by 0.5-6 keV H $^{+}$ and Ne $^{+}$ ion bombardment}}.
\newblock \emph{\bibinfo{journal}{Surface Science}}
  \textbf{\bibinfo{volume}{150}}, \bibinfo{pages}{143--156}
  (\bibinfo{year}{1985}).

\bibitem{Sandford1993}
\bibinfo{author}{{Sandford}, S.~A.} \& \bibinfo{author}{{Allamandola}, L.~J.}
\newblock \bibinfo{title}{{H 2 in Interstellar and Extragalactic Ices: Infrared
  Characteristics, Ultraviolet Production, and Implications}}.
\newblock \emph{\bibinfo{journal}{\apjl}} \textbf{\bibinfo{volume}{409}},
  \bibinfo{pages}{L65} (\bibinfo{year}{1993}).

\bibitem{Watanabe2000}
\bibinfo{author}{{Watanabe}, N.}, \bibinfo{author}{{Horii}, T.} \&
  \bibinfo{author}{{Kouchi}, A.}
\newblock \bibinfo{title}{{Measurements of D$_{2}$ Yields from Amorphous
  D$_{2}$O Ice by Ultraviolet Irradiation at 12 K}}.
\newblock \emph{\bibinfo{journal}{\apj}} \textbf{\bibinfo{volume}{541}},
  \bibinfo{pages}{772--778} (\bibinfo{year}{2000}).

\bibitem{Grieves2005}
\bibinfo{author}{{Grieves}, G.~A.} \& \bibinfo{author}{{Orlando}, T.~M.}
\newblock \bibinfo{title}{{The importance of pores in the electron stimulated
  production of D $_{2}$ and O $_{2}$ in low temperature ice}}.
\newblock \emph{\bibinfo{journal}{Surface Science}}
  \textbf{\bibinfo{volume}{593}}, \bibinfo{pages}{180--186}
  (\bibinfo{year}{2005}).

\bibitem{Zheng2006}
\bibinfo{author}{{Zheng}, W.}, \bibinfo{author}{{Jewitt}, D.} \&
  \bibinfo{author}{{Kaiser}, R.~I.}
\newblock \bibinfo{title}{{Formation of Hydrogen, Oxygen, and Hydrogen Peroxide
  in Electron-irradiated Crystalline Water Ice}}.
\newblock \emph{\bibinfo{journal}{\apj}} \textbf{\bibinfo{volume}{639}},
  \bibinfo{pages}{534--548} (\bibinfo{year}{2006}).

\bibitem{Zheng2006b}
\bibinfo{author}{{Zheng}, W.}, \bibinfo{author}{{Jewitt}, D.} \&
  \bibinfo{author}{{Kaiser}, R.~I.}
\newblock \bibinfo{title}{{Temperature Dependence of the Formation of Hydrogen,
  Oxygen, and Hydrogen Peroxide in Electron-Irradiated Crystalline Water Ice}}.
\newblock \emph{\bibinfo{journal}{\apj}} \textbf{\bibinfo{volume}{648}},
  \bibinfo{pages}{753--761} (\bibinfo{year}{2006}).

\bibitem{Zheng2007}
\bibinfo{author}{{Zheng}, W.}, \bibinfo{author}{{Jewitt}, D.} \&
  \bibinfo{author}{{Kaiser}, R.~I.}
\newblock \bibinfo{title}{{Electron irradiation of crystalline and amorphous D
  $_{2}$O ice}}.
\newblock \emph{\bibinfo{journal}{Chemical Physics Letters}}
  \textbf{\bibinfo{volume}{435}}, \bibinfo{pages}{289--294}
  (\bibinfo{year}{2007}).

\bibitem{Mashchenko2019}
\bibinfo{author}{{Mashchenko}, S.}
\newblock \bibinfo{title}{{Modelling the light curve of `Oumuamua: evidence for
  torque and disc-like shape}}.
\newblock \emph{\bibinfo{journal}{\mnras}} \textbf{\bibinfo{volume}{489}},
  \bibinfo{pages}{3003--3021} (\bibinfo{year}{2019}).

\bibitem{Mamajek2017}
\bibinfo{author}{{Mamajek}, E.}
\newblock \bibinfo{title}{{Kinematics of the Interstellar Vagabond
  1I/{\textquoteleft}Oumuamua (A/2017 U1)}}.
\newblock \emph{\bibinfo{journal}{Research Notes of the American Astronomical
  Society}} \textbf{\bibinfo{volume}{1}}, \bibinfo{pages}{21}
  (\bibinfo{year}{2017}).

\bibitem{Hallatt2020}
\bibinfo{author}{{Hallatt}, T.} \& \bibinfo{author}{{Wiegert}, P.}
\newblock \bibinfo{title}{{The Dynamics of Interstellar Asteroids and Comets
  within the Galaxy: An Assessment of Local Candidate Source Regions for
  1I/'Oumuamua and 2I/Borisov}}.
\newblock \emph{\bibinfo{journal}{\aj}} \textbf{\bibinfo{volume}{159}},
  \bibinfo{pages}{147} (\bibinfo{year}{2020}).

\bibitem{Park2018}
\bibinfo{author}{{Park}, R.~S.}, \bibinfo{author}{{Pisano}, D.~J.},
  \bibinfo{author}{{Lazio}, T. J.~W.}, \bibinfo{author}{{Chodas}, P.~W.} \&
  \bibinfo{author}{{Naidu}, S.~P.}
\newblock \bibinfo{title}{{Search for OH 18 cm Radio Emission from 1I/2017 U1
  with the Green Bank Telescope}}.
\newblock \emph{\bibinfo{journal}{\aj}} \textbf{\bibinfo{volume}{155}},
  \bibinfo{pages}{185} (\bibinfo{year}{2018}).
\newblock \eprint{1803.10187}.

\bibitem{Maggiolo2020}
\bibinfo{author}{{Maggiolo}, R.} \emph{et~al.}
\newblock \bibinfo{title}{{The Effect of Cosmic Rays on Cometary Nuclei. II.
  Impact on Ice Composition and Structure}}.
\newblock \emph{\bibinfo{journal}{\apj}} \textbf{\bibinfo{volume}{901}},
  \bibinfo{pages}{136} (\bibinfo{year}{2020}).

\bibitem{Gronoff2020}
\bibinfo{author}{{Gronoff}, G.} \emph{et~al.}
\newblock \bibinfo{title}{{The Effect of Cosmic Rays on Cometary Nuclei. I.
  Dose Deposition}}.
\newblock \emph{\bibinfo{journal}{\apj}} \textbf{\bibinfo{volume}{890}},
  \bibinfo{pages}{89} (\bibinfo{year}{2020}).

\bibitem{Derenne2010}
\bibinfo{author}{{Derenne}, S.} \& \bibinfo{author}{{Robert}, F.}
\newblock \bibinfo{title}{{Model of molecular structure of the insoluble
  organic matter isolated from Murchison meteorite}}.
\newblock \emph{\bibinfo{journal}{\maps}} \textbf{\bibinfo{volume}{45}},
  \bibinfo{pages}{1461--1475} (\bibinfo{year}{2010}).

\bibitem{Rubin2019}
\bibinfo{author}{{Rubin}, M.} \emph{et~al.}
\newblock \bibinfo{title}{{Elemental and molecular abundances in comet
  67P/Churyumov-Gerasimenko}}.
\newblock \emph{\bibinfo{journal}{\mnras}} \textbf{\bibinfo{volume}{489}},
  \bibinfo{pages}{594--607} (\bibinfo{year}{2019}).

\bibitem{Laufer1987}
\bibinfo{author}{{Laufer}, D.}, \bibinfo{author}{{Kochavi}, E.} \&
  \bibinfo{author}{{Bar-Nun}, A.}
\newblock \bibinfo{title}{{Structure and dynamics of amorphous water ice}}.
\newblock \emph{\bibinfo{journal}{\prb}} \textbf{\bibinfo{volume}{36}},
  \bibinfo{pages}{9219--9227} (\bibinfo{year}{1987}).

\bibitem{Bar-Nun1988}
\bibinfo{author}{{Bar-Nun}, A.} \& \bibinfo{author}{{Prialnik}, D.}
\newblock \bibinfo{title}{{The Possible Formation of a Hydrogen Coma around
  Comets at Large Heliocentric Distances}}.
\newblock \emph{\bibinfo{journal}{\apjl}} \textbf{\bibinfo{volume}{324}},
  \bibinfo{pages}{L31} (\bibinfo{year}{1988}).

\bibitem{Gundlach2012}
\bibinfo{author}{{Gundlach}, B.} \& \bibinfo{author}{{Blum}, J.}
\newblock \bibinfo{title}{{Outgassing of icy bodies in the Solar System - II:
  Heat transport in dry, porous surface dust layers}}.
\newblock \emph{\bibinfo{journal}{\icarus}} \textbf{\bibinfo{volume}{219}},
  \bibinfo{pages}{618--629} (\bibinfo{year}{2012}).

\bibitem{Prialnik2022}
\bibinfo{author}{{Prialnik}, D.} \& \bibinfo{author}{{Jewitt}, D.}
\newblock \bibinfo{title}{{Amorphous ice in comets: evidence and
  consequences}}.
\newblock \emph{\bibinfo{journal}{arXiv e-prints}}
  \bibinfo{pages}{arXiv:2209.05907} (\bibinfo{year}{2022}).

\bibitem{Sekanina2019}
\bibinfo{author}{{Sekanina}, Z.}
\newblock \bibinfo{title}{{Outgassing As Trigger of 1I/`Oumuamua's
  Nongravitational Acceleration: Could This Hypothesis Work at All?}}
\newblock \emph{\bibinfo{journal}{arXiv e-prints}}
  \bibinfo{pages}{arXiv:1905.00935} (\bibinfo{year}{2019}).

\bibitem{Seligman2018}
\bibinfo{author}{{Seligman}, D.} \& \bibinfo{author}{{Laughlin}, G.}
\newblock \bibinfo{title}{{The Feasibility and Benefits of In Situ Exploration
  of {\textquoteleft}Oumuamua-like Objects}}.
\newblock \emph{\bibinfo{journal}{\aj}} \textbf{\bibinfo{volume}{155}},
  \bibinfo{pages}{217} (\bibinfo{year}{2018}).

\bibitem{Fink2009}
\bibinfo{author}{{Fink}, U.}
\newblock \bibinfo{title}{{A taxonomic survey of comet composition 1985-2004
  using CCD spectroscopy}}.
\newblock \emph{\bibinfo{journal}{\icarus}} \textbf{\bibinfo{volume}{201}},
  \bibinfo{pages}{311--334} (\bibinfo{year}{2009}).

\bibitem{Newburn1985}
\bibinfo{author}{{Newburn}, R.~L.} \& \bibinfo{author}{{Spinrad}, H.}
\newblock \bibinfo{title}{{Spectrophotometry of seventeen comets. II - The
  continuum}}.
\newblock \emph{\bibinfo{journal}{\aj}} \textbf{\bibinfo{volume}{90}},
  \bibinfo{pages}{2591--2608} (\bibinfo{year}{1985}).

\bibitem{stern1990}
\bibinfo{author}{Stern, S.}
\newblock \bibinfo{title}{Ism-induced erosion and gas-dynamical drag in the
  oort cloud}.
\newblock \emph{\bibinfo{journal}{Icarus}} \textbf{\bibinfo{volume}{84}},
  \bibinfo{pages}{447--466} (\bibinfo{year}{1990}).

\bibitem{Meltzer1949}
\bibinfo{author}{{Meltzer}, B.}
\newblock \bibinfo{title}{{Shadow Area of Convex Bodies}}.
\newblock \emph{\bibinfo{journal}{\nat}} \textbf{\bibinfo{volume}{163}},
  \bibinfo{pages}{220} (\bibinfo{year}{1949}).

\bibitem{britt2006}
\bibinfo{author}{{Britt}, D.~T.}, \bibinfo{author}{{Consolmagno}, G.~J.} \&
  \bibinfo{author}{{Merline}, W.~J.}
\newblock \bibinfo{title}{{Small Body Density and Porosity: New Data, New
  Insights}}.
\newblock In \bibinfo{editor}{{Mackwell}, S.} \& \bibinfo{editor}{{Stansbery},
  E.} (eds.) \emph{\bibinfo{booktitle}{37th Annual Lunar and Planetary Science
  Conference}}, Lunar and Planetary Science Conference, \bibinfo{pages}{2214}
  (\bibinfo{year}{2006}).

\bibitem{Steckloff2021}
\bibinfo{author}{{Steckloff}, J.~K.} \emph{et~al.}
\newblock \bibinfo{title}{{The sublimative evolution of (486958) Arrokoth}}.
\newblock \emph{\bibinfo{journal}{\icarus}} \textbf{\bibinfo{volume}{356}},
  \bibinfo{pages}{113998} (\bibinfo{year}{2021}).

\end{thebibliography}

\begin{addendum}
  \item[Acknowledgements] We thank Dave Jewitt, Mike Brown, Davide Farnocchia, Karen Meech, Dong Lai, Alessandro Morbidelli,  Luke Kelley, Jonathan Lunine and Samantha Trumbo for useful conversations and suggestions. We thank the scientific editor, Leslie Sage, and the two anonymous reviewers for insightful comments and constructive suggestions that greatly strengthened the scientific content of this manuscript. DZS acknowledges financial support from the National Science Foundation  Grant No. AST-17152, NASA Grant No. 80NSSC19K0444 and NASA Contract  NNX17AL71A from the NASA Goddard Spaceflight Center.  J.B.B. acknowledges support from NASA through the NASA Hubble Fellowship grant HST-HF2-51429.001-A awarded by the Space Telescope Science Institute, which is operated by the Association of Universities for Research in Astronomy, Incorporated, under NASA contract NAS5-26555.
  \item[Author contributions] J.~B.~B.~performed the H$_2$ budget modeling and led the writing of the manuscript.  D.~Z.~S.~performed the thermal modeling and contributed to the manuscript text.
  \item[Data and Code availability] The datasets generated and analysed during the current study are available on GitHub at https://github.com/bergnerjb/Oumuamua\_H2.  
  All code used for this manuscript is available on GitHub at https://github.com/bergnerjb/Oumuamua\_H2.
  
 \item[Competing Interests] The authors declare that they have no competing financial interests.
 \item[Correspondence] Correspondence and requests for materials
should be addressed to J.~B.~B.~(jbergner@berkeley.edu).
\end{addendum}

\end{document}